\documentclass[runningheads]{llncs}
\usepackage[T1]{fontenc}
\usepackage{graphicx}
\usepackage[hidelinks]{hyperref}
\usepackage[capitalise,noabbrev]{cleveref}
\usepackage{microtype}
\usepackage{enumitem}

\begin{document}
\title{Chat-Based Support Alone May Not Be Enough: Comparing Conversational and Embedded LLM Feedback for Mathematical Proof Learning}
\titlerunning{Chat-Based Support Alone May Not Be Enough}

\author{Eason Chen\inst{1} \and
Sophia Judicke\inst{1} \and
Kayla Beigh\inst{1} \and
Xinyi Tang\inst{1} \and
Yumo Wang\inst{1} \and
Mingyu Yuan\inst{1} \and
Zimo Xiao\inst{1} \and
Chuangji Li\inst{1} \and
Shizhuo Li\inst{1} \and
Reed Luttmer\inst{1} \and
Shreya Singh\inst{1} \and
Maria Yampolsky\inst{1} \and
Naman Parikh\inst{1} \and
Yvonne Zhao\inst{1} \and
Meiyi Chen\inst{1} \and
Cheng Huang\inst{1} \and
Anishka Mohanty\inst{1} \and
Gregory Johnson\inst{1} \and
John Mackey\inst{1} \and
Jionghao Lin\inst{2} \and
Ken Koedinger\inst{1}}
\authorrunning{Chen et al.}
\institute{Carnegie Mellon University, Pittsburgh, PA, USA \and
The University of Hong Kong, Hong Kong, China}
\maketitle
\begin{abstract}
We evaluate \textit{GPTutor}, an LLM-powered tutoring system for an undergraduate discrete mathematics course, integrating a structured proof-review tool and a conversational chatbot. In a staggered-access study with 148 students, earlier access was associated with higher homework performance, but this benefit did not transfer to exam scores. A serial mediation analysis reveals distinct pathways for the two components: lower self-efficacy predicted higher usage of both tools, but only chatbot usage constituted a negative associative pathway to subsequent exam performance, while proof-review usage showed no detectable independent association. These findings suggest that conversational LLM support alone may not reliably support transfer to independent assessment, whereas structured, work-anchored feedback showed no detectable negative association with learning outcomes.
\end{abstract}


\section{Introduction}

Proof writing is a core learning objective in undergraduate discrete mathematics \cite{thurston1995proof,schoenfeld2010series,stewart2019student}. Many students struggle to construct valid arguments without timely feedback \cite{weber2001student,stewart2019student}, and learning from proofs depends on active sense-making, especially self-explaining why each step follows from earlier statements \cite{chi1994eliciting,hodds2014self}. Proof learning may therefore be sensitive not only to whether help is available, but also to whether it preserves students' own reasoning rather than outsourcing planning or verification.

Large language models (LLMs) now make it possible to provide scalable, on-demand help \cite{wardat2023chatgpt,chen2023gptutor,chen2024gptutor,cao2024llm,lin2024mufin,han2024improving}. Yet conversational LLM support can reduce student-generated reasoning, particularly when learners rely on the system for planning and checking \cite{chen2025ai,chen2025identifying,zhai2024effects}. This suggests that access alone may not be enough: outcomes may depend on how students use the support and which interface designs they rely on.

We developed \textit{GPTutor}, an LLM-powered tutoring system deployed in an undergraduate discrete mathematics course, integrating a conversational chatbot for math questions and a structured proof-review interface providing localized feedback on students' written proof attempts. Using a staggered-access deployment with 148 students, interaction logs, course performance records, and a baseline self-efficacy survey, we address two core questions:

\begin{enumerate}[nosep]
    \item Does earlier access to GPTutor improve homework and exam performance?
    \item Do the chatbot and proof-review components constitute distinct associative pathways between learner characteristics and exam outcomes?
\end{enumerate}

\section{Related Work}

\noindent\textbf{LLM-Based Educational Support and Overreliance.}
Prior studies report benefits of LLM-based learning support, including increased engagement and self-regulated learning \cite{wardat2023chatgpt,wu2024promoting,chen2024systematic,li2024bringing}, but also highlight risks of overreliance and reduced generative processing \cite{chen2025ai,zhai2024effects,schemmer2023appropriate,frieder2024mathematical}. Most existing studies focus on short-term or controlled settings \cite{wu2024promoting}, offering limited insight into how reliance patterns develop during sustained, real-world use.

\noindent\textbf{AI Support for Proof Writing.}
Proof writing poses unique challenges due to its open-ended structure, reliance on conceptual understanding, and need for logical justification \cite{weber2001student,fujita2018learners}. Traditional intelligent tutoring systems have shown success in well-structured domains \cite{aleven2002effective}, but scaling to proof-based tasks requires extensive domain modeling. LLMs offer flexible feedback without predefined solution paths \cite{khan2024khanmigo}, including AI-driven feedback for open-ended responses \cite{zhao2025slideitright,armfield2025avalon}, but empirical evidence for complex proof learning remains limited \cite{liu2025understanding}.

\noindent\textbf{Self-Efficacy and Help-Seeking.}
Self-efficacy plays a central role in shaping engagement with challenging tasks, influencing persistence, strategy use, and help-seeking \cite{bandura1997self,hackett1985role}. Students with lower self-efficacy are more likely to rely heavily on external assistance \cite{schunk1991self,zimmerman2000self}, and in AI-supported environments, this may amplify risks of unproductive reliance. Our study addresses this gap by analyzing how reliance on conversational versus structured LLM support relates to self-efficacy and exam performance in an undergraduate proof-based course.

\section{System Design}

\textit{GPTutor} provides two core functionalities to support students' engagement with open-ended proof-writing homework, designed to scaffold reflection on logical structure while preserving opportunities for productive struggle. Full implementation details are at appendix \url{https://github.com/EasonC13/AIED26-Chatbots}.

\begin{figure}[t!]
    \centering
    \includegraphics[width=1\linewidth]{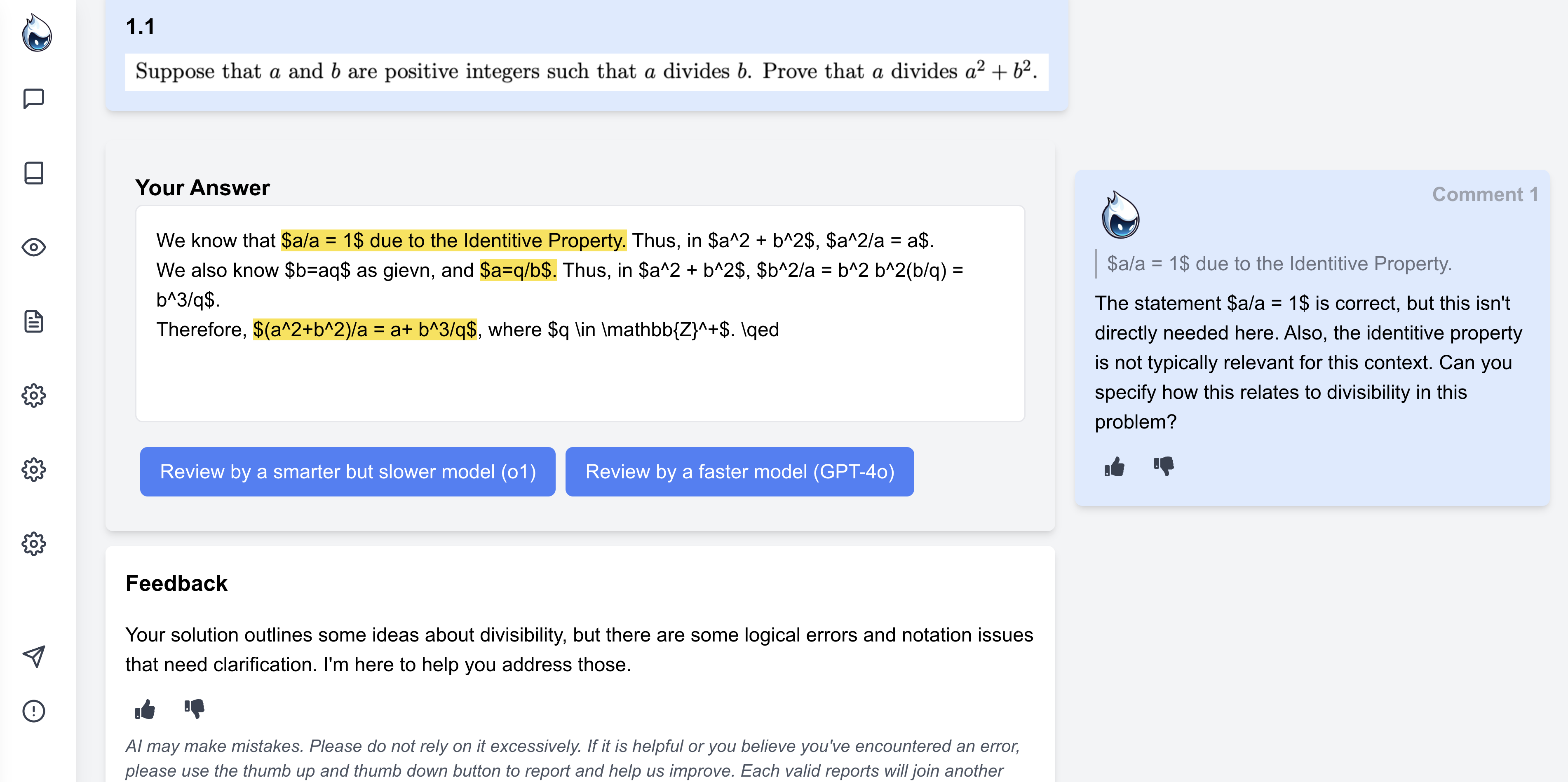}
    \caption{Proof-Review-GPTutor interface: high-level comments and localized highlights on specific proof segments.}
    \label{fig:homework_reviewer}
\end{figure}

\noindent\textbf{Proof-Review-GPTutor.}
The proof-review component provides both high-level feedback on the overall proof and localized comments highlighting specific problematic segments (\cref{fig:homework_reviewer}) \cite{lin2024can}. Drawing on research about how explanation presentation shapes AI reliance \cite{vasconcelos2023explanations}, visual highlighting makes feedback localized and interpretable. Instructor-provided reference solutions are supplied via in-context learning. To reduce answer extraction, the system supports only structured feedback and does not allow open-ended dialogue. Students must first produce a proof attempt before receiving feedback, anchoring the interaction in their own work.

\begin{figure}[t!]
    \centering
    \includegraphics[width=1\linewidth]{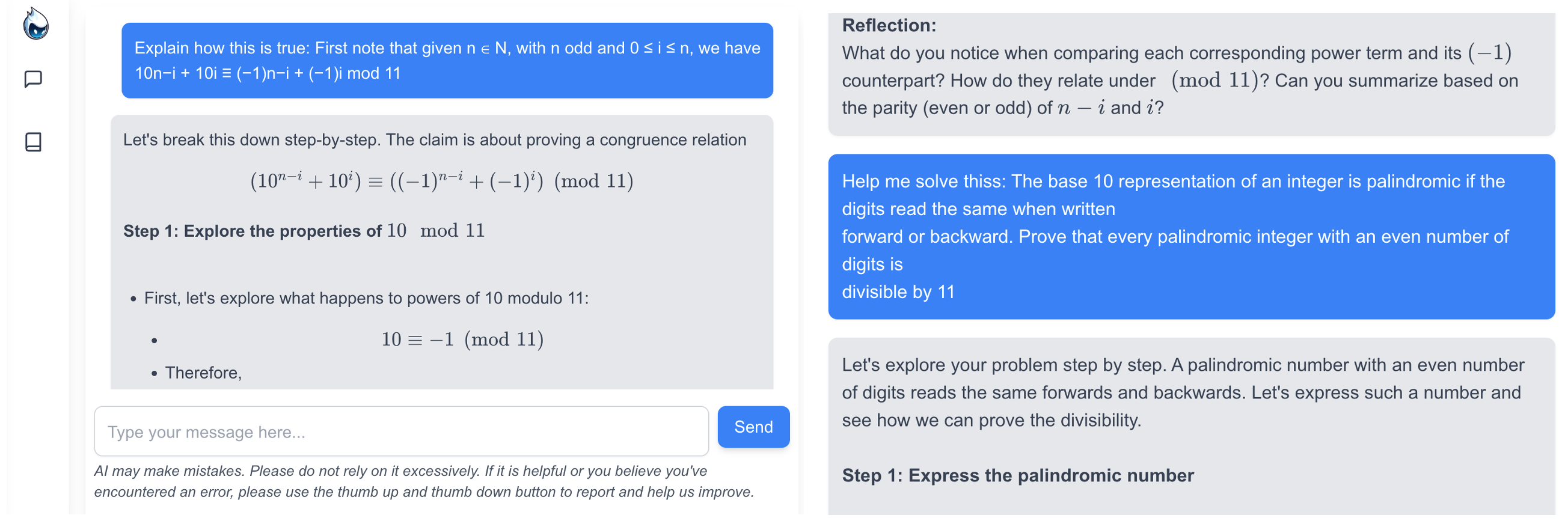}
    \caption{Chatbot interface. Left: step-by-step guidance with a reflection prompt. Right: subsequent conversation that the student bypasses scaffolding to seek direct solutions.}
    \label{fig:chatbot_example}
\end{figure}

\noindent\textbf{AI Chatbot.}
The chatbot (GPT-4o) supports conversational, math-focused dialogue (\cref{fig:chatbot_example}). A system prompt constrains conversation to mathematics and instructs the model to use Socratic questioning and partial hints rather than direct answers. However, some students exhibited answer-seeking and escalation behaviors, bypassing scaffolding to seek solutions. An audit of 80 randomly sampled sessions confirmed 100\% adherence to the answer-withholding prompt (two independent coders, full agreement).

\section{Methods}

\subsection{Participants and Study Design}

This Institutional Review Board (IRB)-approved study deployed \textit{GPTutor} in an undergraduate discrete mathematics course (308 enrolled) at a large US research university. Students were invited to consent and complete a baseline survey after Midterm~1. After consent, 155 participants were randomly assigned: the experimental group ($n=77$) received GPTutor access after Midterm~1, while the control group ($n=78$) received access after Midterm~2. This staggered-access design provided descriptive comparisons of performance during periods when access differed, while ensuring all participants ultimately received access. Homework assignments were submitted in \LaTeX{}, and students were encouraged (but not required) to paste drafts into the Proof-Review-GPTutor prior to submission. Course policies did not restrict students' use of other external resources outside exams. After attrition, 73 experimental and 72 control students completed the study. Students were not permitted to use AI tools during exams.

\subsection{Measures}

\noindent\textbf{Academic Performance.} Performance outcomes included scores from ten homework assignments and three midterm exams (Midterm~1--3), graded by course staff using standard procedures.

\noindent\textbf{System Interaction Logs.} \textit{GPTutor} logged all interactions with both components. We recorded timestamps and counts of proof-review requests and chatbot messages, aggregated by exam intervals to align reliance with subsequent assessments. For example, analyses predicting Midterm~3 used interactions between Midterm~2 and Midterm~3.

\noindent\textbf{Self-Efficacy.} At baseline (after Midterm~1), participants completed a self-efficacy scale adapted from validated instruments \cite{pintrich1990motivational}. Internal consistency was high (Cronbach's $\alpha = 0.91$).

\subsection{Analysis Strategy}

\noindent\textbf{Access-level comparisons.} Mixed-effects regressions on homework scores (random intercept for student, controlling for prior homework) and ordinary least squares (OLS) regressions on exam scores controlling for Midterm~1.

\noindent\textbf{Serial mediation analysis.} We estimated a serial mediation model reflecting temporal ordering: baseline self-efficacy $\rightarrow$ Midterm~2 $\rightarrow$ component usage (Midterm~2$\rightarrow$Midterm~3 interval) $\rightarrow$ Midterm~3 performance, with direct paths retained. Bootstrapped confidence intervals (5{,}000 resamples; \cite{preacher2004spss}) tested indirect effects. Mediation analyses focus on the Midterm~2$\rightarrow$Midterm~3 interval, when all participants had access, maximizing sample size and comparability.

\section{Results}

Unless otherwise noted, reliance-based analyses focus on the \textbf{Midterm~2$\rightarrow$Midterm~3} interval, when both conditions had access to GPTutor. Additional evaluation details are available in the appendix repository: \url{https://github.com/EasonC13/AIED26-Chatbots} and in the earlier preprint version of this work \cite{chen2025generative}.

\subsection{Access improves homework but does not transfer to exams}

\begin{figure}[t!]
    \centering
    \includegraphics[width=0.85\linewidth]{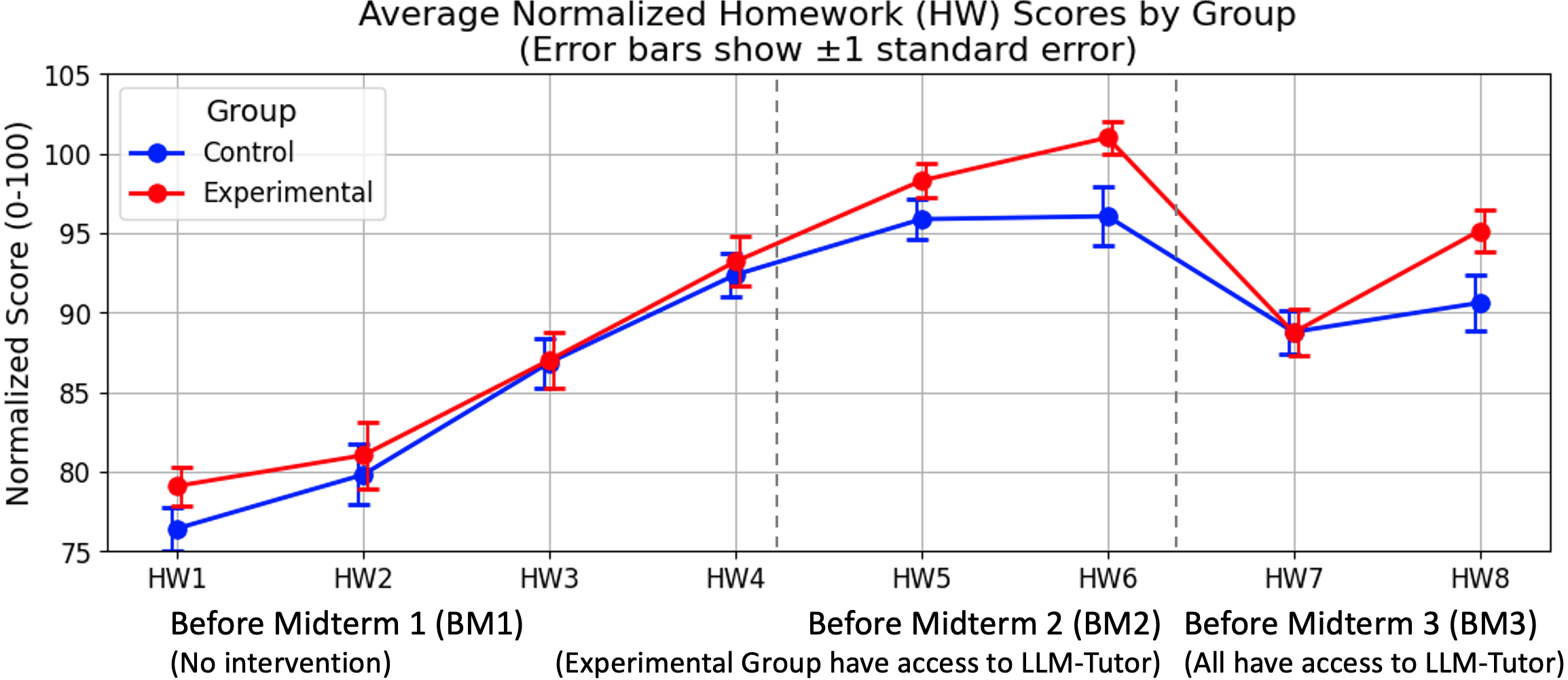}
    \caption{Normalized \textbf{homework} scores by condition. During the period when only the experimental group had access, they achieved higher scores on average.}
    \label{fig:jmackey_homework_grade}
\end{figure}

As shown in \cref{fig:jmackey_homework_grade}, we fit a mixed-effects regression on \textbf{HW5--HW6} scores (the interval when only the experimental group had access), including a random intercept for student and controlling for prior homework performance (\textbf{HW1--HW4} average). After adjustment, earlier access was associated with higher \textbf{HW5--HW6} scores ($B=2.71$, $SE=1.22$, 95\% CI $[0.32, 5.09]$, $p=.026$), corresponding to a small-to-medium standardized effect ($\beta=0.23$). Baseline homework performance strongly predicted \textbf{HW5--HW6} outcomes ($B=0.60$, $SE=0.05$, $p<.001$), and the intraclass correlation indicated that 24\% of variance reflected stable between-student differences. Once the control group also gained access (\textbf{HW7--HW8}), the difference disappeared ($B=1.06$, $p=.432$), consistent with the staggered-access design.

In contrast, OLS regressions predicting \textbf{Midterm~2} and \textbf{Midterm~3} from condition (controlling for \textbf{Midterm~1}) showed no significant group differences (\textbf{Midterm~2}: $B=-2.32$, $p=.134$; \textbf{Midterm~3}: $B=2.46$, $p=.283$). Taken together, GPTutor improved homework performance for whichever group had access, but this benefit did not transfer to independent exam assessments.

\subsection{Serial mediation: distinct pathways for chatbot and proof-review}

\begin{figure}[t!]
    \centering
    \includegraphics[width=1\linewidth]{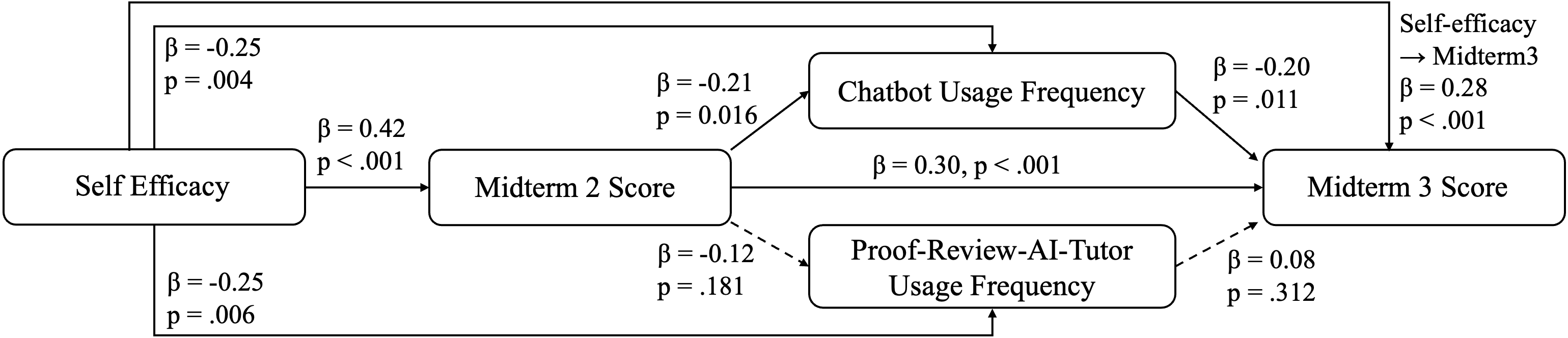}
    \caption{Serial mediation model. Standardized coefficients ($\beta$) and $p$-values on arrows. Usage measured during \textbf{Midterm~2$\rightarrow$Midterm~3}.}
    \label{fig:serial_mediation_model}
\end{figure}

Before estimating the mediation model, we confirmed that usage of both components reflected help-seeking patterns. Self-efficacy was negatively associated with both chatbot usage frequency ($r=-.34$, $p<.001$) and proof-review usage frequency ($r=-.30$, $p<.001$). Similarly, students with lower \textbf{Midterm~2} scores used both tools more frequently during the subsequent interval (chatbot: $r=-.31$, $p<.001$; proof-review: $r=-.22$, $p=.008$).

The serial mediation model (\cref{fig:serial_mediation_model}) revealed distinct patterns. Lower self-efficacy predicted lower \textbf{Midterm~2} performance ($B=3.95$, $p<.001$) and higher usage of both tools (chatbot: $B=-20.52$, $p=.004$; proof-review: $B=-3.59$, $p=.006$). \textbf{Midterm~2} uniquely predicted chatbot usage beyond self-efficacy ($B=-1.82$, $p=.016$), but not proof-review usage ($B=-0.18$, $p=.181$). Critically, chatbot usage was negatively associated with \textbf{Midterm~3} ($B=-0.031$, $p=.011$), whereas proof-review usage was not ($B=0.068$, $p=.312$). Direct effects of self-efficacy ($B=3.74$, $p<.001$) and \textbf{Midterm~2} ($B=0.42$, $p<.001$) on \textbf{Midterm~3} remained substantial.

Bootstrapped indirect effects confirmed a partial mediation pathway from self-efficacy to \textbf{Midterm~3} via chatbot usage (indirect $=0.69$, 95\% CI $[0.14, 1.56]$), whereas the proof-review pathway was not significant (indirect $=-0.24$, 95\% CI $[-0.72, 0.12]$). The self-efficacy $\rightarrow$ \textbf{Midterm~2} $\rightarrow$ \textbf{Midterm~3} pathway was also significant (indirect $=1.70$, 95\% CI $[0.76, 2.83]$). The direct effect of self-efficacy on \textbf{Midterm~3} remained substantial ($c'=3.74$, total $c=6.04$).

\section{Discussion}

\noindent\textbf{Access aligns with improved homework but not exam transfer.}
Earlier access to GPTutor coincided with higher homework performance during the period when only the experimental group could use the system, whereas midterm exam performance did not differ by access condition. This contrast suggests a boundary on what LLM-based support influences in this setting. Homework affords iterative drafting and revision, aligning with repeated feedback cycles. Exams, in contrast, require independent proof construction without access to tools. The absence of exam differences therefore points to limited transfer from AI-supported practice to independent assessment performance, consistent with broader concerns about LLM-assisted learning \cite{zhai2024effects}.

\noindent\textbf{Conversational reliance as a risk pathway.}
The serial mediation model reveals that chatbot usage constitutes a measurable pathway linking lower self-efficacy and prior difficulty to poorer exam outcomes. In contrast, proof-review usage reflects help-seeking without detectable outcome associations after controlling for individual differences. This divergence highlights that simply deploying LLMs as a chatbot is not sufficient for proof learning. Interface constraints likely shape whether AI support preserves learner sense-making: the proof-review requires students to first produce a proof attempt and then engages them with localized, structured feedback tied to their own work, potentially reducing opportunities for short-circuiting reasoning. Conversational support, even with pedagogical prompts, can still invite repeated reliance under uncertainty \cite{zhai2024effects,schemmer2023appropriate,vasconcelos2023explanations}.

An analogy to traditional help-seeking contexts clarifies this contrast: chatbot interactions resemble asking someone for an answer, where the open-ended format naturally invites requests for solutions even when students are instructed to engage pedagogically. The proof-review interface, by contrast, resembles submitting a draft for feedback: students must first produce their own work, and the system responds with comments anchored to specific parts of their attempt, constraining the interaction toward instructional feedback rather than direct answers.

These findings should not be interpreted as evidence that chatbot use causes lower scores; a plausible alternative is that already-struggling students are both more likely to seek conversational support and more likely to perform worse. However, the persistence of associations after controlling for prior performance suggests that conversational traces can serve as useful signals for identifying students who may need additional support or different scaffolds \cite{zhai2024effects,schemmer2023appropriate}.

One notable tension is that higher chatbot usage was negatively associated with exam performance at the individual level, yet the experimental group as a whole did not perform worse on midterms. A possible interpretation is that negative associations with chatbot reliance were offset by benefits from the proof-review component. Future work comparing access to proof-review alone versus combined access could clarify whether work-anchored feedback yields more positive effects when not bundled with conversational support.

\noindent\textbf{Limitations and future work.}
Although participants were randomly assigned to staggered access, course policies permitted other external resources, so our comparisons reflect associations under naturalistic conditions rather than isolated causal effects. We do not measure time-on-task, displacement effects (e.g., whether chatbot use replaced independent practice or other resources), or within-student change in self-efficacy, which was measured once and assumed stable. Detailed behavioral coding of answer-seeking and escalation patterns, outcome associations controlling for behavioral labels, and three-group comparisons among non-users and different user types remain for future work.

Future work could complement usage-frequency analyses with behavioral coding of student-AI interactions to distinguish productive help-seeking, answer-seeking, and other usage patterns that may explain why different interface designs yield different learning outcomes. Moreover, future work should combine richer process measures (e.g., self-reported study diaries) with interventions that redirect conversational reliance toward generative processing and evaluate transfer on independent assessments. Our findings suggest that LLM integration in proof learning should move beyond generic chat interfaces toward designs that anchor feedback in student work and actively regulate reliance, potentially incorporating richer feedback modalities such as animated pedagogical agents \cite{chen2025vtutor_aied} or real-time tutoring support systems \cite{chen2025vtutor_las}.

\bibliographystyle{splncs04}
\bibliography{reference}

@inproceedings{liu2025understanding,
  title={Understanding student engagement with large language model-powered course assistants},
  author={Liu, Chang and Hoang, Loc and Stolman, Andrew and Kizilcec, Rene F and Wu, Bo},
  full_author={Liu, Chang and Hoang, Kien and Stolman, Noah and Kizilcec, René F and Wu, Ting-Ting},
  booktitle={International Conference on Artificial Intelligence in Education},
  pages={3--10},
  year={2025},
  organization={Springer}
}

@article{aleven2002effective,
  title={An effective metacognitive strategy: Learning by doing and explaining with a computer-based cognitive tutor},
  author={Aleven, Vincent AWMM and Koedinger, Kenneth R},
  full_author={Aleven, Vincent AWMM and Koedinger, Kenneth R},
  journal={Cognitive science},
  volume={26},
  number={2},
  pages={147--179},
  year={2002},
  publisher={Wiley Online Library}
}

@article{zhai2024effects,
  title={The effects of over-reliance on AI dialogue systems on students' cognitive abilities: a systematic review},
  author={Zhai, Chunpeng and Wibowo, Santoso and Li, Lily D},
  full_author={Zhai, Chunpeng and Wibowo, Santoso and Li, Li},
  journal={Smart Learning Environments},
  volume={11},
  number={1},
  pages={28},
  year={2024},
  publisher={Springer}
}

@article{chen2024systematic,
  title={A systematic review on prompt engineering in large language models for k-12 stem education},
  author={Chen, Eason and Wang, Danyang and Xu, Luyi and Cao, Chen and Fang, Xiao and Lin, Jionghao},
  full_author={Chen, Eason and Wang, Danyang and Li, Jeffrey and Huang, Scarlett and Tang, Xinyi and Lin, Jionghao},
  journal={arXiv preprint arXiv:2410.11123},
  year={2024}
}

@article{fujita2018learners,
  title={Learners’ use of domain-specific computer-based feedback to overcome logical circularity in deductive proving in geometry},
  author={Fujita, Taro and Jones, Keith and Miyazaki, Mikio},
  full_author={Fujita, Taro and Jones, Keith and Miyazaki, Masami},
  journal={ZDM},
  volume={50},
  number={4},
  pages={699--713},
  year={2018},
  publisher={Springer}
}

@misc{khan2024khanmigo,
  title={Khanmigo},
  author={Khan, Sal},
  year={2024}
}

@article{wu2024promoting,
  title={Promoting self-regulation progress and knowledge construction in blended learning via ChatGPT-based learning aid},
  author={Wu, Ting-Ting and Lee, Hsin-Yu and Li, Pin-Hui and Huang, Chia-Nan and Huang, Yueh-Min},
  full_author={Wu, Ting-Ting and Lee, Meng-Han and Li, Chia-Chi and Huang, Yu-Min and Huang, Yueh-Min},
  journal={Journal of Educational Computing Research},
  volume={61},
  number={8},
  pages={3--31},
  year={2024},
  publisher={SAGE Publications Sage CA: Los Angeles, CA}
}

@article{vasconcelos2023explanations,
  title={Explanations can reduce overreliance on ai systems during decision-making},
  author={Vasconcelos, Helena and J{\"o}rke, Matthew and others},
  full_author={Vasconcelos, Helena and Jörke, Matthew and Kale, Alex and Krüger, Andreas and Stumpe, Martin C and Braga, Diogo},
  journal={Proceedings of the ACM on Human-Computer Interaction},
  volume={7},
  number={CSCW1},
  pages={1--38},
  year={2023},
  publisher={ACM New York, NY, USA}
}

@article{wardat2023chatgpt,
  title={ChatGPT: A revolutionary tool for teaching and learning mathematics},
  author={Wardat, Yousef and Tashtoush, Mohammad A and AlAli, Rommel and Jarrah, Adeeb M},
  full_author={Wardat, Yousef and Tashtoush, Mohammad A and AlAli, Rommel and Jarrah, Adeeb M},
  journal={Eurasia Journal of Mathematics, Science and Technology Education},
  volume={19},
  number={7},
  pages={em2286},
  year={2023},
  publisher={Modestum}
}

@article{stewart2019student,
  title={Student perspectives on proof in linear algebra},
  author={Stewart, Sepideh and Thomas, Michael OJ},
  full_author={Stewart, Sepideh and Thomas, Michael OJ},
  journal={ZDM},
  volume={51},
  number={7},
  pages={1069--1082},
  year={2019},
  publisher={Springer}
}

@article{frieder2024mathematical,
  title={Mathematical capabilities of chatgpt},
  author={Frieder, Simon and Pinchetti, Luca and others},
  full_author={Frieder, Simon and Pinchetti, Luca and Griffiths, Ryan-Rhys and Salvatori, Tommaso and Lukasiewicz, Thomas and Petersen, Philipp and Chevalier, Alexis and Berner, Julius},
  journal={Advances in neural information processing systems},
  volume={36},
  year={2024}
}

@article{li2024bringing,
  title={Bringing generative AI to adaptive learning in education},
  author={Li, Hang and Xu, Tianlong and others},
  full_author={Li, Hang and Xu, Weiqi and Zhang, Jiani and Chen, Eason and Lin, Jionghao and Gurung, Ashish and Han, Zifei FeiFei and Thomas, Danielle R and Koedinger, Kenneth R},
  journal={arXiv preprint arXiv:2402.14601},
  year={2024}
}

@article{schunk1991self,
  title={Self-efficacy and academic motivation},
  author={Schunk, Dale H},
  journal={Educational psychologist},
  volume={26},
  number={3-4},
  pages={207--231},
  year={1991},
  publisher={Taylor \& Francis}
}

@article{hackett1985role,
  title={Role of mathematics self-efficacy in the choice of math-related majors of college women and men: A path analysis.},
  author={Hackett, Gail},
  journal={Journal of counseling psychology},
  volume={32},
  number={1},
  pages={47},
  year={1985},
  publisher={American Psychological Association}
}

@article{pintrich1990motivational,
  title={Motivational and self-regulated learning components of classroom academic performance.},
  author={Pintrich, Paul R and others},
  full_author={Pintrich, Paul R and De Groot, Elisabeth V},
  journal={Journal of educational psychology},
  volume={82},
  number={1},
  pages={33},
  year={1990},
  publisher={American Psychological Association}
}

@article{zimmerman2000self,
  title={Self-efficacy: An essential motive to learn},
  author={Zimmerman, Barry J},
  journal={Contemporary educational psychology},
  volume={25},
  number={1},
  pages={82--91},
  year={2000},
  publisher={Elsevier}
}

@book{bandura1997self,
  title={Self-efficacy: The exercise of control},
  author={Bandura, Albert},
  year={1997},
  publisher={Macmillan}
}

@article{thurston1995proof,
  title={On proof and progress in mathematics},
  author={Thurston, William P},
  journal={For the learning of mathematics},
  volume={15},
  number={1},
  pages={29--37},
  year={1995},
  publisher={JSTOR}
}

@article{schoenfeld2010series,
  title={Series editor's foreword: The soul of mathematics In DA Stylianou, ML Blanton, \& EJ Knuth},
  author={Schoenfeld, AH},
  journal={Teaching and learning proof across the grades: A K-16 perspective},
  year={2010}
}

@article{preacher2004spss,
  title={SPSS and SAS procedures for estimating indirect effects in simple mediation models},
  author={Preacher, Kristopher J and Hayes, Andrew F},
  full_author={Preacher, Kristopher J and Hayes, Andrew F},
  journal={Behavior research methods, instruments, \& computers},
  volume={36},
  pages={717--731},
  year={2004},
  publisher={Springer}
}

@article{chi1994eliciting,
  title={Eliciting self-explanations improves understanding},
  author={Chi, Michelene T. H. and de Leeuw, Nicholas and Chiu, Mei-Hung and LaVancher, Christian},
  full_author={Chi, Michelene T. H. and de Leeuw, Nicholas and Chiu, Mei-Hung and LaVancher, Christian},
  journal={Cognitive science},
  volume={18},
  number={3},
  pages={439--477},
  year={1994},
  publisher={Elsevier}
}

@article{weber2001student,
  title={Student difficulty in constructing proofs: The need for strategic knowledge},
  author={Weber, Keith},
  journal={Educational studies in mathematics},
  volume={48},
  number={1},
  pages={101--119},
  year={2001},
  publisher={Springer}
}

@inproceedings{schemmer2023appropriate,
  title={Appropriate reliance on AI advice: Conceptualization and the effect of explanations},
  author={Schemmer, Max and Kuehl, Niklas and Benz, Carina and Bartos, Andrea and Satzger, Gerhard},
  full_author={Schemmer, Max and Kühl, Niklas and Benz, Claudia and Bartos, Michael and Satzger, Gerhard},
  booktitle={Proceedings of the 28th International Conference on Intelligent User Interfaces},
  pages={410--422},
  year={2023}
}

@article{hodds2014self,
  title={Self-explanation training improves proof comprehension},
  author={Hodds, Mark and Alcock, Lara and Inglis, Matthew},
  full_author={Hodds, Mark and Alcock, Lara and Inglis, Matthew},
  journal={Journal for Research in Mathematics Education},
  volume={45},
  number={1},
  pages={62--101},
  year={2014},
  publisher={National Council of Teachers of Mathematics}
}

@article{han2024improving,
  title={Improving assessment of tutoring practices using retrieval-augmented generation},
  author={Han, Zifei FeiFei and Lin, Jionghao and Gurung, Ashish and others},
  full_author={Han, Zifei FeiFei and Lin, Jionghao and Gurung, Ashish and Thomas, Danielle R and Chen, Eason and Borchers, Conrad and Gupta, Shivang and Koedinger, Kenneth R},
  journal={arXiv preprint arXiv:2402.14594},
  year={2024}
}

@inproceedings{cao2024llm,
  title={Llm-generated personalized analogies to foster ai literacy in adult novices},
  author={Cao, Cassie Chen and Zoe, FANG and Lydia, Y CAO and Jionghao, LIN},
  full_author={Cao, Cassie Chen and Chen, Eason and Li, Jeffrey and Huang, Scarlett and Tang, Xinyi and Lin, Jionghao},
  booktitle={International Conference on Computers in Education},
  year={2024}
}

@article{chen2025ai,
  title={AI Knows Best? The Paradox of Expertise, AI-Reliance, and Performance in Educational Tutoring Decision-Making Tasks},
  author={Chen, Eason and Li, Jeffrey and Huang, Scarlett and Tang, Xinyi and Lin, Jionghao and Carvalho, Paulo F and Koedinger, Kenneth R},
  full_author={Chen, Eason and Li, Jeffrey and Huang, Scarlett and Tang, Xinyi and Lin, Jionghao and Carvalho, Paulo F and Koedinger, Kenneth R},
  journal={arXiv preprint arXiv:2509.16772},
  year={2025}
}

@inproceedings{chen2025identifying,
  title={Identifying Effective Praise in Tutoring: Large Language Models with Transparent Explanations},
  author={Chen, Eason and Li, Jeffrey and Huang, Scarlett and Tang, Xinyi and Lin, Jionghao and others},
  full_author={Chen, Eason and Li, Jeffrey and Huang, Scarlett and Tang, Xinyi and Lin, Jionghao and Carvalho, Paulo and Koedinger, Kenneth},
  booktitle={International Conference on Artificial Intelligence in Education},
  pages={157--163},
  year={2025},
  organization={Springer}
}

@inproceedings{zhao2025slideitright,
  title={Slideitright: using AI to find relevant slides and provide feedback for open-ended questions},
  author={Zhao, Chloe Qianhui and Cao, Jie and Chen, Eason and Koedinger, Kenneth R and Lin, Jionghao},
  full_author={Zhao, Chloe Qianhui and Cao, Jie and Chen, Eason and Lin, Jionghao and Koedinger, Kenneth R},
  booktitle={International Conference on Artificial Intelligence in Education},
  pages={378--392},
  year={2025},
  organization={Springer}
}

@inproceedings{chen2025vtutor_aied,
  title={VTutor: An Animated Pedagogical Agent SDK that Provide Real Time Multi-Model Feedback},
  author={Chen, Eason and Lin, Chenyu and Huang, Yu-Kai and Tang, Xinyi and Xi, Aprille and Lin, Jionghao and Koedinger, Kenneth},
  full_author={Chen, Eason and Lin, Chenyu and Lin, Jionghao and Tang, Xinyi and Gao, Ruijie and Wang, Isabel and Koedinger, Kenneth R},
  booktitle={International Conference on Artificial Intelligence in Education},
  pages={152--159},
  year={2025},
  organization={Springer}
}

@inproceedings{chen2025vtutor_las,
  title={Vtutor for high-impact tutoring at scale: managing engagement and real-time multi-screen monitoring with p2p connections},
  author={Chen, Eason and Tang, Xinyi and others},
  full_author={Chen, Eason and Tang, Xinyi and Lin, Chenyu and Wang, Isabel and Lin, Jionghao and Gao, Ruijie and Parikh, Naman and Koedinger, Kenneth R},
  booktitle={Proceedings of the Twelfth ACM Conference on Learning@ Scale},
  pages={320--324},
  year={2025}
}

@inproceedings{chen2023gptutor,
  title={GPTutor: a ChatGPT-powered programming tool for code explanation},
  author={Chen, Eason and others},
  full_author={Chen, Eason and Huang, Ray and Chen, Han-Shin and Tseng, Yuen-Hsien and Li, Liang-Yi},
  booktitle={International Conference on Artificial Intelligence in Education},
  year={2023},
  organization={Springer}
}

@inproceedings{armfield2025avalon,
  title={Avalon: a human-in-the-loop LLM grading system with instructor calibration and student self-assessment},
  author={Armfield, Derek and Chen, Eason and others},
  full_author={Armfield, Derek and Chen, Eason and Li, Jeffrey and Huang, Scarlett and Tang, Xinyi and Lin, Jionghao and Koedinger, Kenneth R},
  booktitle={International Conference on Artificial Intelligence in Education},
  pages={111--118},
  year={2025}
}

@inproceedings{chen2024gptutor,
  title={GPTutor: Great personalized tutor with large language models for personalized learning content generation},
  author={Chen, Eason and Lee, Jia-En and Lin, Jionghao and Koedinger, Kenneth R},
  full_author={Chen, Eason and Lee, Jia-En and Lin, Jionghao and Koedinger, Kenneth R},
  booktitle={Proceedings of the Eleventh ACM Conference on Learning@ Scale},
  pages={539--541},
  year={2024}
}

@inproceedings{lin2024mufin,
  title={MuFIN: a framework for automating multimodal feedback generation using generative artificial intelligence},
  author={Lin, Jionghao and Chen, Eason and Gurung, Ashish and Koedinger, Kenneth R},
  full_author={Lin, Jionghao and Chen, Eason and Gurung, Ashish and Koedinger, Kenneth R},
  booktitle={Proceedings of the Eleventh ACM Conference on Learning@ Scale},
  pages={550--552},
  year={2024}
}

@inproceedings{lin2024can,
  title={How can i improve? using gpt to highlight the desired and undesired parts of open-ended responses},
  author={Lin, Jionghao and Chen, Eason and others},
  full_author={Lin, Jionghao and Chen, Eason and Han, Zifei FeiFei and Gurung, Ashish and Thomas, Danielle R and Cao, Cassie Chen and Borchers, Conrad and Koedinger, Kenneth R},
  booktitle={Proceedings of the 17th International Conference on Educational Data Mining},
  pages={236--250},
  year={2024}
}

@article{chen2025generative,
  title={Generative AI alone may not be enough: Evaluating AI Support for Learning Mathematical Proof},
  author={Chen, Eason and Judicke, Sophia and others},
  full_author={Chen, Eason and Judicke, Sophia and Beigh, Kayla and Tang, Xinyi and Xiao, Zimo and Li, Chuangji and Li, Shizhuo and Luttmer, Reed and Singh, Shreya and Yampolsky, Maria and Parikh, Naman and Zhao, Yvonne and Chen, Meiyi and Huang, Cheng and Mohanty, Anishka and Johnson, Gregory and Mackey, John and Lin, Jionghao and Koedinger, Kenneth R},
  journal={arXiv preprint arXiv:2509.16778},
  year={2025}
}

\end{document}